\documentclass[journal,10pt]{IEEEtran}
\IEEEoverridecommandlockouts
\usepackage{cite}
\usepackage{soul}
\usepackage[numbers,sort&compress]{natbib}
\usepackage{amsmath,amssymb,amsfonts}
\usepackage{algorithm,algpseudocode}
\usepackage{graphicx}
\floatname{algorithm}{Algorithm}
\usepackage{titlesec}
\makeatletter
\newcommand*{\rom}[1]{\expandafter\@slowromancap\romannumeral #1@}
\makeatother
\usepackage{hyperref}
\usepackage{caption}
\usepackage{fixltx2e}
\usepackage{tabularx}
\usepackage{makecell}
\hypersetup{
colorlinks=false,    
linkbordercolor={1 1 1}, 
citebordercolor={1 1 1}, 
urlbordercolor={1 1 1}   
}

\usepackage[table]{xcolor}
\usepackage{xcolor}
\usepackage{float}

\usepackage{tikz}

\usepackage{listings}
\definecolor{codegreen}{rgb}{0,0.6,0}
\definecolor{codegray}{rgb}{0.5,0.5,0.5}
\definecolor{codepurple}{rgb}{0.58,0,0.82}
\definecolor{backcolour}{rgb}{0.92,0.92,0.92}
\usepackage{textcomp}
\newcommand{\RNum}[1]{\uppercase\expandafter{\romannumeral #1\relax}}
\usepackage{xcolor}
\usepackage{listings}
\def\BibTeX{{\rm B\kern-.05em{\sc i\kern-.025em b}\kern-.08em
	T\kern-.1667em\lower.7ex\hbox{E}\kern-.125emX}}
	
\begin{document}

\title{ChIRAAG: ChatGPT Informed Rapid and Automated Assertion Generation}

\author{\IEEEauthorblockN{Bhabesh Mali*, Karthik Maddala*, Vatsal Gupta*, Sweeya Reddy*, Chandan Karfa*, Ramesh Karri\dag}\\
	\IEEEauthorblockA{*Indian Institute of Technology Guwahati, India\\\dag New York University, USA\\
		\ *\{m.bhabesh, k.maddala, g.vatsal, r.poddutoori, ckarfa\}@iitg.ac.in, \dag rkarri@nyu.edu}}

\maketitle

\begin{abstract}
System Verilog Assertion (SVA) formulation- a critical yet complex task is a prerequisite in the \textcolor{black}{Assertion Based Verification (ABV) process}. Traditionally, SVA formulation involves expert-driven interpretation of specifications, which is time-consuming and prone to human error. Recently, LLM-informed automatic assertion generation is gaining interest. We designed a novel framework called ChIRAAG, based on OpenAI GPT4, to generate SVA from natural language specifications \textcolor{black}{of a design}. ChIRAAG constitutes the systematic breakdown of design specifications into a standardized format, further generating assertions from formatted specifications using LLM. Furthermore, we \textcolor{black}{used few test cases} to validate the LLM-generated assertions. Automatic feedback of log messages from the simulation tool to the LLM ensures that the framework can generate correct SVAs. \textcolor{black}{In our experiments,} only 27\%  of LLM-generated raw assertions had errors, which was rectified in few iterations based on the simulation log. Our results on OpenTitan designs show that LLMs can streamline and assist engineers in the assertion generation process, reshaping verification workflows.
\end{abstract}


\section{Introduction}
Pre-silicon bug detection is a crucial step in designing functionally correct hardware. As per the functional verification study conducted by Wilson Research Group in 2020 \cite{Foster_2021}, over 50\% of the overall development resources and expenditures in Application-Specific Integrated Circuit (ASIC) and Field-Programmable Gate Array (FPGA) systems were allocated to the verification process. Assertion Based Verification (ABV) validates the correct design implementation by integrating safety and liveness properties, expressed as assertions, throughout the development process. \textcolor{black}{Safety property states ``something bad does not happen". While liveness property states ``something good will eventually happen" \cite{dasgupta2006roadmap}.} However, as the scale of design and verification requirements increases, so does the number of assertions, which can reach hundreds. \textcolor{black}{Identifying and writing manually all assertions of a design is a non-trivial task.}

Large Language Models (LLMs) have evolved to assist humans in solving complex problems. LLMs \textcolor{black}{can also help in} ABV by automatically generating the assertions from the natural language specification \textcolor{black}{of a design}. The existing methods have relied on generating SVA from Verilog code \cite{orenesvera2023using}, comments, and assertion-based examples \cite{kande2023llmassisted}. None of them could generate SVA directly from the specifications. This is the motivation for this study to formulate a novel framework that reduces human effort and time in \textcolor{black}{SVA formulation}. 
This work aims to analyze how LLMs can assist \textcolor{black}{SVA generation semi-automatically} from the natural language specifications. In our framework, manual intervention is needed when there is a potential bug in the implementation \textcolor{black}{or the LLM cannot fix the issue in the generated assertion(s) based on the simulation log.} We seek to answer two key questions through this work: \textit{RQ1: How good are LLMs in generating System Verilog Assertion (SVA) from natural language specification?} \textit{RQ2: How functionally correct are the generated assertions within the design and verification framework?}. 

Our contributions are fourfold.
\begin{itemize}
	\item A framework to study the use of LLM in assertion generation for ABV.
	\item \textcolor{black}{The standardization of the format for specification and a prompting method to generate SVA.}
	\item An analysis of the performance of LLMs in generating SVA for OpenTitan designs \cite{Opentitan}.
	\item Recommendations of LLMs for assertion generation.
\end{itemize}

\textcolor{black}{The rest of the paper is organized as follows: Section \rom{2} discusses the related works. Section \rom{3} describes our proposed framework. Section \rom{4} shows a case study on a particular design. The experimentation and performance results of our proposed approach are illustrated in Section \rom{5}. Section \rom{6} concludes the paper with future directions.}
\section{Background and related works}

LLMs are Transformer Language models where the construction of the weight parameters involves training models using an extensive volume of text data acquired from diverse sources such as various websites, web-books, and PDFs. 
LLMs have leading-edge capabilities in diverse tasks, including source code generation \cite{chen2021evaluating}, automated testing \cite{deng2023large}, and program repair \cite{xia2022less}. As in~\cite{orenesvera2023using}, the authors showcased the capability of LLM to generate assertions solely from RTL code without additional specifications. Despite initial syntactic and semantic errors, the authors observed that, with proper guidance, LLMs consistently produce correct and complete assertions. The work in \cite{kande2023llmassisted} uses comments about the security assertions to generate assertions. They assessed the effectiveness of LLMs in accelerating property verification. They generated 75,600 assertions across diverse conditions and varying levels of context and found that LLMs can generate correct assertions at an average rate of 4.53\%. The work in \cite{witharana2023automated} uses LLMs in security assertion generation. They perform formal property verification to identify vulnerabilities before developing the Design Under Test (DUT).

In contrast, we have exclusively used the design specification in natural language to generate SVA instead of directly providing RTL code for assertion generation to the LLM. Since the objective of assertions is to identify the bug in the implementation, it is not desirable to obtain the assertions from implementation. Moreover, creating assertions from a design specification poses a significant challenge in the absence of a dedicated and specialized framework. \textcolor{black}{We have also observed that the recent works are non-iterative, i.e., once the assertions are generated, the frameworks don't try to refine any errors, such as syntactic or simulation errors, generated by LLM. }Therefore, we proposed a novel framework, ChIRAAG, that formats the input specification and gives it to the LLM and further 
reviews, analyzes, and mends the LLM-generated assertions, if required. We have mainly focused on generating functional assertions rather than generating security assertions. None of the above-discussed works generate assertions using only the specification.

\section{ChIRAAG: Our SVA Generation Framework}

The automatic assertion generation with ChIRAAG involves two stages: \textcolor{black}{(i) formatting of specifications and (ii) automatic assertion generation.} \textbf{Algorithm} \ref{Algorithm1} shows the steps of our proposed ChIRAAG framework. The stages are briefly described below:

\subsection{Formatting of specification}
The formatting of design specifications includes representing the design context with specific labels, informing about its general definitions, signals, functional parameters, etc. The formatting involves extracting the important information from the design specification, $\mathcal{D}$, and later converting the extracted information, $UnFmt$, into a Javascript Object Notation (JSON) format, $\mathcal{J}$\textcolor{black}{, using the functions \textit{Extraction} and \textit{JSON}, as mentioned in Line 6 of \textbf{Algorithm \ref{Algorithm1}}. The $JSON$ function comprises several sub-functions as shown in Table \ref{table:formatting} to produce the formatted specification. }

\begin{table}[tbh]
	\caption{Formatting of specification}
	\label{table:formatting}
	\centering
	\begin{tabular}{|l|l|}
		\hline
		\textbf{Label} & \textbf{Description} \\
		\hline
		\texttt{Introduction} & ${\mathcal{J}_{int}} \gets JSON_{int}$($UnFmt$) \\
		\texttt{System\_overview} & ${\mathcal{J}_{s\_ov}} \gets JSON_{s\_ov}$ ( $UnFmt$) \\
		\texttt{Definitions} & $\mathcal{J}_{df} \gets JSON_{df}$ ($UnFmt$) \\
		\texttt{Parameters} & $\mathcal{J}_{par} \gets JSON_{par}$ ($UnFmt$) \\
		\texttt{Functional\_requirements} & $\mathcal{J}_{fn\_r} \gets JSON_{fn\_r}$ ($UnFmt$) \\
		\texttt{Timing\_requirements} & $\mathcal{J}_{tm\_r} \gets JSON_{tm\_r}$ ($UnFmt$) \\
		\texttt{Extra\_info} & $\mathcal{J}_{ext} \gets JSON_{ext}$  ($UnFmt$) \\
		\hline
	\end{tabular}
\end{table}

\subsection{Automatic Assertion Generation}
\textcolor{black}{In the next step, ChIRAAG generates} SVA using the JSON formatted specifications. The formatted specifications of the designs are provided as input to the LLM through our framework by making a call to the OpenAI Application Programming Interface (API). This generates raw SVA as stated in Line 10 of \textbf{Algorithm \ref{Algorithm1}}. The generated assertions are further verified for correctness in a simulation environment using testcases. This generates \textcolor{black}{simulation logs that are then parsed using the `log\_parser' function, producing log message, $\mathcal{W}_{log}$, as shown in Line 13 of \textbf{Algorithm \ref{Algorithm1}}.} This step gives rise to four possible cases that can eventually occur. They are 

\textit{(a) No Error Message:} When the generated assertion, $A_n$, passes all the test cases, i.e., no error message in $\mathcal{W}_{log}$, then $A_n$ is taken as correct output assertions suite. 

\textit{(b) Syntax Error:} Some LLM-generated assertions have Syntax Errors (SE). \textcolor{black}{The LLM is provided the log message, $\mathcal{W}_{log}$, automatically. Further, the LLM is prompted to refine the assertion.} LLM is able to correct the SE within a few iterations. Line 17 of \textbf{Algorithm \ref{Algorithm1}} refine the assertion based on the error message.

\textit{(c) Simulation Failure:} The assertions generated by LLM might fail in the simulation run because of specific errors. Two such simulation errors we have encountered are timing errors and missing signals. \textcolor{black}{Line 20 of \textbf{Algorithm \ref{Algorithm1}} refine the assertion based on the error message}.

\textit{(d) Implementation Bug: }It might be possible that even the generated assertions are correct, but the implementation might be buggy. In such a scenario, the testcase fails, giving wrong outputs. In this case, manual inspection of design implementation is to be done as shown in Line 23 of \textbf{Algorithm \ref{Algorithm1}}.

The assertions may not pass the testcases in a single run. Therefore, we iteratively run the error-checking process multiple times. Line 11 of \textbf{Algorithm \ref{Algorithm1}} shows that the while loop runs until $n$ is less than the upper bound $\mathcal{T}$. If $n$ gets equal to $\mathcal{T}$, and there are still error messages in $\mathcal{W}_{log}$, \textcolor{black}{we manually investigate and resolve the potential issue which LLM cannot fix automatically and restart the assertion generation process as stated in Lines 29-33.}

\begin{figure}[hbt!]
	\centering 
	\captionsetup{justification=centering}
	\label{fig_11}
	\includegraphics[width=.87\linewidth]{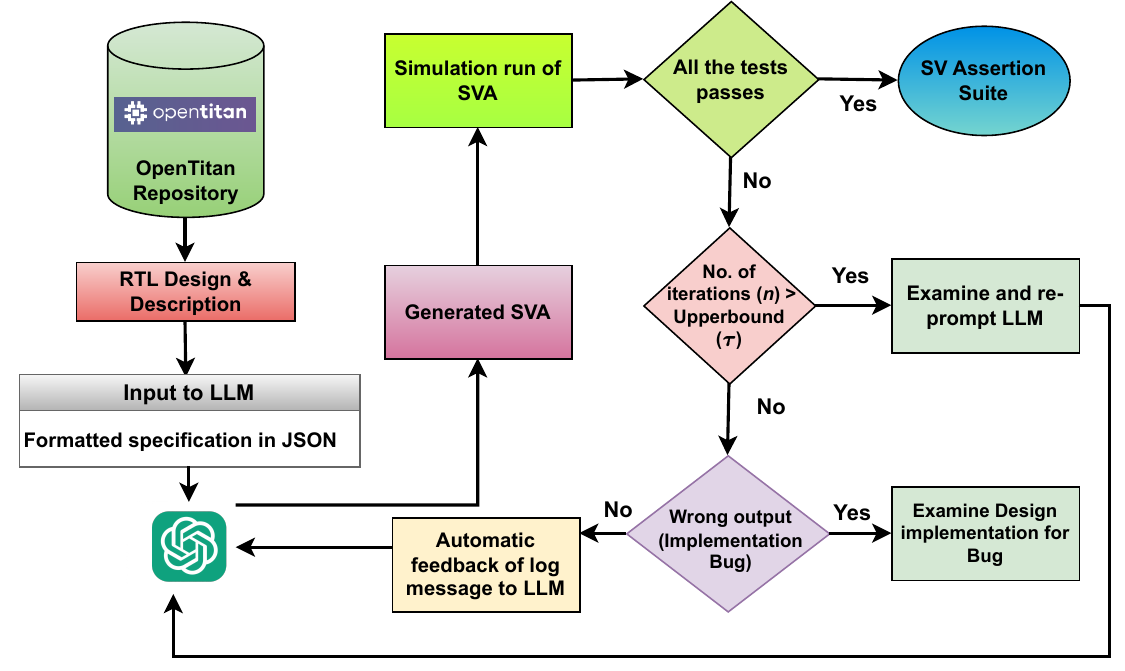}
	\caption{ChIRAAG-SVA Generation Framework}
	\label{fig: flow_diagram}
\end{figure}

The proposed ChIRAAG framework is illustrated in the Fig. \ref{fig: flow_diagram}. \textcolor{black}{ We use the design only to check the syntax and semantics of the generated assertions. No other refinement of assertions is done using the design. In this process, a correctly generated assertion can fix a bug (if any) in a design as well.}

\begin{algorithm}
	\begin{small}
		\caption{ChIRAAG: SVA Generation Framework}
		\label{Algorithm1}
		\begin{algorithmic}[1]
			\State \textbf{Input:} 
			\State \hspace*{\algorithmicindent} $\mathcal{D}$: Design specification document, $\mathcal{L}$: Log file of simulation tool, $\mathcal{T}$: Upper bound on the number of LLM interaction prompts
			\State \textbf{Output:} 
			\State \hspace*{\algorithmicindent} $A_n$: Suite of System Verilog Assertions

			\Procedure{Specification Extraction and Formatting}{}
			\State $UnFmt \gets Extraction (\mathcal{D})$, $\mathcal{J} \gets \text{JSON}(UnFmt)$
			\EndProcedure
			
			\Procedure{LLM Interaction for SVA Generation}{}
			\State Initialize $n \gets 0 \text{ and } \mathcal{L} \gets \{\}$
			\State $A_n \gets \text{LLM}(\mathcal{J}, \mathcal{L})$
			\While{$n < \mathcal{T}$}
			\State Use VCS to check the generated $A_n$ on testcases
			\State $\mathcal{W}_{log} \gets \text{log\_parser}(\mathcal{L})$
			\If{no errors in $\mathcal{W}_{log}$}
			\State \textbf{break}
			\ElsIf{syntactic errors in $\mathcal{W}_{log}$}
			\State $\overline{A}_n \gets $ Rectify syntactic errors
			\State $A_n \gets \overline{A_n}$
			\ElsIf{timing or missing signal error in $\mathcal{W}_{log}$}
			\State $\overline{A}_n \gets$ Rectify simulation errors
			\State $A_n \gets \overline{A_n}$
			\ElsIf{testcase failure message in $\mathcal{W}_{log}$}
			\State Manual inspection of design implementation for bug
			\State \textbf{break}
			\EndIf
			\State $n \gets n + 1$
			\EndWhile
			\EndProcedure
			
			\Procedure{Design Implementation Examination}{}
			\If{$n == \mathcal{T}$ and error message in $\mathcal{W}_{log}$}
			\State Examine design implementation and restart the process
			\EndIf
			\EndProcedure
			
			
		\end{algorithmic}
	\end{small}
\end{algorithm}

\section{Case Study: RV Timer}
\begin{figure}[htb]
	\centering
	\includegraphics[width=0.9\linewidth]{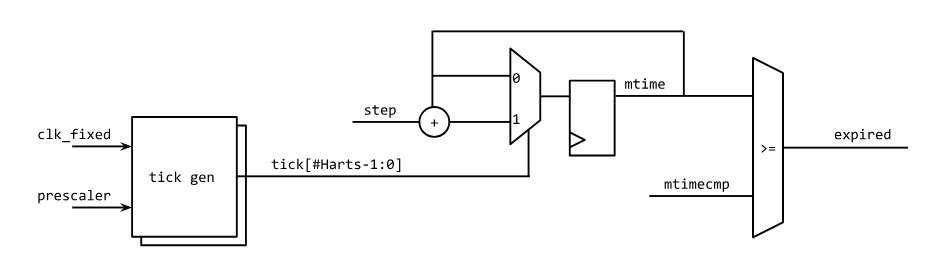}
	\caption{Block Diagram of \texttt{RV Timer}}
	\label{fig:block-rv-timer}
\end{figure}

\textcolor{black}{\texttt{RV Timer} \cite{Opentitan} is a design module that is intended for use by processors to monitor the current time relative to the system reset or power-on.} The block diagram of the \texttt{RV Timer} is shown in Fig. \ref{fig:block-rv-timer}

The unstructured design specification of \texttt{RV Timer} is initially transformed to structured JSON format and presented as input to the LLM to generate the SVA. The LLM initially generated ten functional assertions. We further evaluated the correctness of the raw assertions using testcases and VCS simulator. Seven out of ten assertions were correct (i.e., there are no syntax or simulation errors) in their raw format. This shows the potential of our proposed framework ChIRAAG. The remaining three raw assertions were rectified automatically using the steps of \textbf{Algorithm \ref{Algorithm1}}. This process generated an extra assertion, which satisfies the design aspects, increasing the count of generated assertions to eleven. Some of the initial assertions of \texttt{RV Timer} generated by the LLM are:

\begin{lstlisting}
	
	(*@\textbf{Assertion 1:}@*) 
	// Assertion to check if tick_count resets to 0 on reset
	property p_reset_tick_count;
	@(posedge clk_i) (!rst_ni) |-> (tick_count == 12'h0);
	end property
	assert property (p_reset_tick_count);
	
	
	(*@\textbf{Assertion 2:}@*)
	// Assertion to check if tick_count increments correctly
	property p_tick_count_increment;
	@(posedge clk_i) (active && (tick_count < prescaler)) |-> (tick_count == $past(tick_count) + 1);
	end property
	assert property (p_tick_count_increment);
\end{lstlisting}


The \textbf{Assertion 1} is a correct raw assertion that passes all the tests. It ensures that `\texttt{tick\_count}' resets to 0 when reset is asserted, i.e., when `\texttt{rst\_ni}' or the active low reset signal is not active. The \textbf{Assertion 2}, we observed that a missing signal prevents passing all the test cases. The LLM understood the missing signal \texttt{rst\_ni} from the simulation log message and included it in the generated assertion in the next iteration. Line 20 of \textbf{Algorithm} \ref{Algorithm1} is called upon to rectify the \textbf{Assertion 2}. The LLM generated the refined assertion and modified it into two distinct assertions {\textbf{Modified Assertion $\textbf{2a}$}} and {\textbf{Modified Assertion $\textbf{2b}$}}.

\begin{lstlisting}
	(*@\textbf{Modified Assertion 2a:}@*) 
	// Assertion for tick_count increment
	property tick_count_increment;
	@(posedge clk_i)
	(rst_ni && active && (tick_count < prescaler)) |=> (tick_count == $past(tick_count) + 1);
	endproperty
	assert property (tick_count_increment);
	
	(*@\textbf{Modified Assertion 2b:}@*) 
	// Assertion for tick_count reset on reset deassertion
	property tick_count_reset_on_reset_deassert;
	@(posedge clk_i)
	(!$past(rst_ni, 1) && rst_ni) -> (tick_count == 0);
	endproperty
	assert property (tick_count_reset_on_reset_deassert);
\end{lstlisting}

The \textcolor{black}{\textbf{Modified Assertion $\textbf{2a}$}} ensures that as long as the system is not in reset, is active, and the tick count is below a set threshold, the tick count will increment by 1 in each clock cycle. The \textcolor{black}{\textbf{Modified Assertion $\textbf{2b}$}} verifies that when the system transitions from a reset state to a non-reset state, the tick count is reset to 0, ensuring proper initialization or re-initialization of the system's tick counter. \textcolor{black}{ChIRAAG took twelve iterations to converge because, initially, it has syntax errors like identifier missing and token missing, which are resolved in seven iterations for all the assertions combined, and then simulation errors like timing failure are rectified using further prompting in five more iterations.} 

\section{Experimental Results and Discussion}

Our novel framework, ChIRAAG, was evaluated using designs provided in the OpenTitan \cite{Opentitan} project. We performed our experiments on six designs as shown in Table \ref{table:performance}.

We have utilized OpenAI's GPT-4 turbo \cite{openai2023gpt4} LLM as an automatic assertion generation tool for our ChIRAAG framework. This model has a capacity of 1280000 tokens of context window and training data up to December 2023. We collected the design specification and related files for each design and manually structured the data in a specific format with distinctive labels, and corresponding details for each design.

ChIRAAG generated the correct assertion using LLM within a few automatic prompting processes for each design.
Table \ref{table:performance} \textcolor{black}{shows} the performance report \textcolor{black}{for} LLM-based assertion generation. The first column, ``Module'' shows different designs we have taken. The third column, ``LLM Assert.'', shows the number of SV assertions generated by LLM for each design, and the second column, ``OT Assert.'', shows the number of assertions provided in the RTL of OpenTitan. \textcolor{black}{``\#Prompts'' denotes} the number of prompts required to generate the correct assertions \textcolor{black}{that satisfy all the} test cases. \textcolor{black}{``VCS Sim. Time''} shows the time taken by the simulation \textcolor{black}{process}. \textcolor{black}{While the ``SVA Generation Time''} is the initial assertion generation time \textcolor{black}{taken by} our framework.

ChIRAAG successfully generated all of the OpenTitan assertions for each design. Interestingly, ChIRAAG generates more assertions than that was provided in OpenTitan. Importantly, ChIRAAG-generated assertions are important and satisfy some aspects of the design that were missed by the OpenTitan assertions. Our proposed framework, ChIRAAG, generates the assertions for each design in less than 15 seconds. This shows the effectiveness of ChIRAAG in a faster assertion generation process than the traditional method.

\begin{table}[!t]
	\caption{Performance Report of ChIRAAG}
	\label{table:performance}
	\centering
	\begin{tabular}{|l|c|c|c|c|c|}
		\hline
		\textbf{Module} & \makecell{\textbf{OT} \\ \textbf{Assert.}} & \makecell{\textbf{LLM} \\ \textbf{Assert.}} & \makecell{\textbf{\#Prompts} } & \makecell{\textbf{VCS} \\ \textbf{Sim.}\\ \textbf{Time}} & \makecell{\textbf{SVA} \\ \textbf{Generation} \\ \textbf{Time}} \\
		\hline
		RV Timer & 0 &11 & 12 & 80ns & 6.34s\\
		\hline
		PattGen & 0 & 9 & 9 & 110ns & 9.45s \\
		\hline
		GPIO & 0 & 6 & 8 & 190ns & 12.34s \\
		\hline
		ROM Ctrl& 6 & 11 & 14 & 250ns & 14.34s\\
		\hline
		sram\_ctrl & 0 & 14 & 8 & 100ns & 10.23s \\
		\hline
		adc\_ctrl & 5 & 8 & 9 & 460ns & 7.56s \\
		\hline
	\end{tabular}
\end{table}


While ChIRAAG can rectify syntactic and simulation errors \textcolor{black}{using the log of a simulator like VCS}, it can also detect potential bugs in the implementation. To demonstrate such a scenario, we have considered a buggy implementation of a 4-bit full-adder, where \textit{xor} is replaced by an \textit{or} operation. Initially, formatted specifications were fed to ChIRAAG to generate the SVA. ChIRAAG correctly generated two assertions listed as \textcolor{black}{\textbf{Assertion 3}} and \textcolor{black}{\textbf{Assertion 4}}, for checking the carry-out generated and the other for checking the sum bits generated.
\begin{lstlisting}
	(*@\textbf{Assertion 3:}@*) 
	//Assertion to check if the carry-out (C4) is generated correctly for the addition
	property prop_carry_out;
	@(posedge C0)
	C4 |-> ((A[3] & B[3]) | (C3 & (A[3] ^ B[3])));
	endproperty
	assert property (prop_carry_out);
	(*@\textbf{Assertion 4:}@*) 
	//Assertion to check if the sum bits (S) are correct
	property prop_sum_bits;
	@(posedge C0)
	S[0] |-> (A[0] ^ B[0] ^ C0) &&
	S[1] |-> (A[1] ^ B[1] ^ C1) &&
	S[2] |-> (A[2] ^ B[2] ^ C2) &&
	S[3] |-> (A[3] ^ B[3] ^ C3);
	endproperty
	assert property(prop_sum_bits);
\end{lstlisting}

Even though the above assertions are correct, the testcase fails due to the wrong output. After correcting the implementation to \textit{xor} from \textit{or}, the SVA could pass all the tests, This shows that our automated framework can also detect bugs in design implementation. This shows the usefulness of ChIRAAG in generating SVA from natural language specifications. The basic details of the formatted specification, along with the codebase, simulation files, logs, and prompts, are accessible on our GitHub repository \footnote{https://github.com/karthikmaddala/ChIRAAG}.

\section{Conclusion}
 In this work, we have developed ChIRAAG that can generate SVA from \textcolor{black}{the formatted} specification using LLM. Primarily, unstructured design specifications are transformed into a structured manner and given to the LLM as an input prompt to generate an SVA suite. The prompting is crucial in generating correct assertions. \textcolor{black}{The RTL implementation is only used to simulate and identify syntax and simulation errors in generated assertions.} The performance of ChIRAAG on OpenTitan design shows that an LLM is a good starting point to assist an engineer in assertion generation. We noticed only 27\% of the generated SVA require refinement upon iterative prompting.  We performed our experiment with a general-domain LLM; however, we believe a domain-specific LLM for ABV should provide even more accurate assertions in fewer iterations. \textcolor{black}{In our work, we ensure that the generated assertions are syntactically and semantically correct. However, how meaningful and complete these are, in the context of a design needs to be checked by experts.} It is not possible to confirm 100\% functional coverage of design intent by the generated assertions. Our future goal will be to check the capability of LLM to address the ``consistency" and ``completeness" issues in ABV.

\renewcommand*{\bibfont}{\small}
\bibliographystyle{IEEEtran}
\vspace{-0.1 cm}
\bibliography{ref.bib}

\begin{thebibliography}{10}
\providecommand{\url}[1]{#1}
\csname url@samestyle\endcsname
\providecommand{\newblock}{\relax}
\providecommand{\bibinfo}[2]{#2}
\providecommand{\BIBentrySTDinterwordspacing}{\spaceskip=0pt\relax}
\providecommand{\BIBentryALTinterwordstretchfactor}{4}
\providecommand{\BIBentryALTinterwordspacing}{\spaceskip=\fontdimen2\font plus
\BIBentryALTinterwordstretchfactor\fontdimen3\font minus
  \fontdimen4\font\relax}
\providecommand{\BIBforeignlanguage}[2]{{%
\expandafter\ifx\csname l@#1\endcsname\relax
\typeout{** WARNING: IEEEtran.bst: No hyphenation pattern has been}%
\typeout{** loaded for the language `#1'. Using the pattern for}%
\typeout{** the default language instead.}%
\else
\language=\csname l@#1\endcsname
\fi
#2}}
\providecommand{\BIBdecl}{\relax}
\BIBdecl

\bibitem{Foster_2021}
\BIBentryALTinterwordspacing
H.~Foster, ``Part 1: The 2020 wilson research group functional verification
  study,'' Feb 2021. [Online]. Available:
  \url{https://tinyurl.com/verification-foster-group}
\BIBentrySTDinterwordspacing

\bibitem{dasgupta2006roadmap}
P.~Dasgupta, \emph{A roadmap for formal property verification}.\hskip 1em plus
  0.5em minus 0.4em\relax Springer, 2006.

\bibitem{orenesvera2023using}
M.~Orenes-Vera, ``Using llms to facilitate formal verification of rtl,'' 2023.

\bibitem{kande2023llmassisted}
R.~Kande \emph{et~al.}, ``Llm-assisted generation of hardware assertions,''
  2023.

\bibitem{Opentitan}
OpenTitan, ``{OpenTitan Repository},'' \url{https://opentitan.org/}, Visited:
  2023.

\bibitem{chen2021evaluating}
M.~Chen \emph{et~al.}, ``Evaluating large language models trained on code,''
  \emph{arXiv preprint arXiv:2107.03374}, 2021.

\bibitem{deng2023large}
Y.~Deng \emph{et~al.}, ``Large language models are edge-case fuzzers: Testing
  deep learning libraries via fuzzgpt,'' \emph{arXiv preprint
  arXiv:2304.02014}, 2023.

\bibitem{xia2022less}
C.~S. Xia, ``Less training, more repairing please: revisiting automated program
  repair via zero-shot learning,'' in \emph{Proceedings of ACM Joint European
  Software Engineering Conference and Symposium on Foundations of Software
  Engineering}, 2022.

\bibitem{witharana2023automated}
H.~Witharana \emph{et~al.}, ``Automated generation of security assertions for
  rtl models,'' \emph{ACM Journal on Emerging Technologies in Computing
  Systems}, vol.~19, no.~1, pp. 1--27, 2023.

\bibitem{openai2023gpt4}
OpenAI, ``Gpt-4 technical report,'' 2023.

\end{thebibliography}

\end{document}